\documentclass[a4paper,10pt,twoside]{cpc-hepnp}

\usepackage{multicol}
\usepackage{graphicx}
\usepackage{booktabs}
\usepackage{amssymb,bm,mathrsfs,bbm,amscd}
\usepackage[tbtags]{amsmath}
\usepackage{lastpage}

\begin{document}

\fancyhead[co]{\footnotesize GUO Xin-Heng et al: Supernova Neutrinos
Detection On Earth}


\title{Supernova Neutrinos Detection On Earth\thanks{supported in part
by National Natural
Science Foundation of China (10535050, 10675022), the Key Project of
Chinese Ministry of Education (106024) and the Special Grants from
Beijing Normal University. }}

\author{%
      GUO Xin-Heng$^{1;1)}$\email{xhguo@bnu.edu.cn}%
 \quad HUANG Ming-Yang$^{1;2)}$\email{hmy19151905@mail.bnu.edu.cn}%
 \quad YOUNG Bing-Lin$^{2,3;3)}$\email{young@iastate.edu}
 } \maketitle

\address{%
1~(College of Nuclear Science and Technology, Beijing Normal University,
Beijing 100875, China)\\
2~(Department of Physics and Astronomy, Iowa State University, Ames, Iowa
5001, USA)\\
3~(Institute of Theoretical Physics, Chinese Academy of Sciences, Beijing,
China)\\
}

\begin{abstract}
In this paper, we first discuss the detection of supernova neutrino
on Earth. Then we propose a possible method to acquire information
about $\theta_{13}$ smaller than $1.5^\circ$ by detecting the ratio
of the event numbers of different flavor supernova neutrinos. Such
an sensitivity cannot yet be achieved by the Daya Bay reactor
neutrino experiment.
\end{abstract}

\begin{keyword}
supernova, neutrino, collective effects, MSW effects, Earth matter
effects, Daya Bay
\end{keyword}

\begin{pacs}
14.60.Pq, 13.15.+g, 25.30.Pt
\end{pacs}

\begin{multicols}{2}

\section{Introduction}

The supernova (SN) explosion is one of the most spectacular cosmic
events and a source of new physics ideas\cite{SN1,SN2,SN3,SN4,SN5}.
Observable effects of SN neutrinos in underground detectors have
been a subject of intense investigation in astroparticle physics,
both on general grounds and in relation to the SN event like 1987A.
In particular, flavor oscillation in the SN may shed light on the
problem of neutrino masses and mixing by means of the associated
matter effects. Several neutrino laboratories, including the Daya
Bay reactor neutrino underground laboratory\cite{DayaBay} which is
under construction, can be used to detect possible neutrino events
from an SN explosion and serve as the SN Earth Warning System. Hence
theoretical prediction for the detection of SN neutrinos in the Daya
Bay and other neutrino experiments is very desirable.

In the realistic case, when we detect neutrinos from a type II SN
explosion at Daya Bay, there are three effects (the collective
effects arising from neutrino-neutrino
interactions\cite{Collective1,Collective2,Collective3,Collective4,
Collective5,Collective6,Collective7,Collective8,Collective9}, the
well-known Mikheyev-Smirnov-Wolfenstein (MSW)
effects\cite{MSW1,MSW2,MSW3,MSW4}, and Earth matter
effects\cite{EM1,EM2,EM3}) need to be considered. Using the
Landau-Zener formula\cite{Landau1,Landau2}, the expression of the
crossing probability, $P_H$, which is the neutrino jump probability
from mass eigenstate $\nu_1$ to $\nu_3$ at the high resonance region
inside the SN, can be calculated\cite{PH1,PH2,PH3}. With the
relation between $P_H$ and the mixing angle $\theta_{13}$, we can
predict the SN neutrino event numbers $N$ as a function of
$\theta_{13}$. Therefore, we can propose a possible method to
acquire information about small $\theta_{13}$ through the detection
of SN neutrinos\cite{EM3}.

\section{\label{sec:hamckm}Detection of SN neutrinos on Earth}

SN are extremely powerful explosions in the universe which
terminate the life of some stars\cite{SN1,book1,book2,book3}. They
make the catastrophic end of stars more massive than 8 solar
masses leaving behind compact remnants such as neutron stars or
black holes which may be observed. For historical reasons, SN are
divided into two wide categories (type I and type II)
characterized by the absence or presence of hydrogen lines.
However, the most important physical characteristics is the
mechanism that produces the SN, which distinguishes SN of type Ia
from SN of type Ib, Ic, and II. This difference becomes noticeable
in the light spectrum some months after maximum luminosity, when
the ejecta become optically thin and the innermost regions become
visible: the spectrum of SN Ia is dominated by Fe emission lines,
while SN Ib, Ic, and II show O and C emission lines. From the
point of view of neutrino physics, type Ib, Ic, and II SN are much
more important than type Ia SN, simply because they produce a huge
flux of neutrinos and antineutrinos of all flavors.

The type II SN is thought to be generated by the core collapse of
red (or blue as SN1987A) giant star with a mass between about 8-9
and 40-60 solar masses. The total energy release (about
$3\times10^{53}$) is approximately the gravitational binding energy
of the core. It generates intensive neutrinos which take away about
$99\%$ of this total energy. The explosion itself consumes about
1$\%$ of this total energy.
The vast amount of neutrinos are produced in two bursts. In the
first burst which lasts for only a few milliseconds, electron
neutrinos are generated via the electron capture by nuclei
$e^-+N(Z,A)\rightarrow N(Z-1,A)+\nu_e$ and the inverse beta-decay
$e^-+p\rightarrow n+\nu_e$. In the second burst which lasts longer,
neutrinos of all flavors ($\nu_{\alpha}$ and $\bar{\nu}_{\alpha}$
with $\alpha$ being $e$, $\mu$, $\tau$) are produced through the
electron-positron pair annihilation $e^-+e^+\rightarrow
\nu_{\alpha}+\bar{\nu}_{\alpha}$, electron-nucleon bremsstrahlung
$e^{\pm}+ N\rightarrow e^{\pm}+N+\nu_{\alpha}+\bar{\nu}_{\alpha}$,
nucleon-nucleon bremsstrahlung $N+N\rightarrow
N+N+\nu_{\alpha}+\bar{\nu}_{\alpha}$, plasmon decay
$\gamma\rightarrow \nu_{\alpha}+\bar{\nu}_{\alpha}$, and
photoannihilation $\gamma+e^{\pm}\rightarrow
e^{\pm}+\nu_{\alpha}+\bar{\nu}_{\alpha}$\cite{book1,book2,book3}.
When SN neutrinos of a definite flavor are produced they are
approximately in an effective mass eigenstate due to the extremely
high matter density environment. While they propagate outward to the
surface of the SN they could experience collective
effects\cite{Collective1,Collective2,Collective3,Collective4,Collective5,
Collective6,Collective7,Collective8,Collective9} and MSW
effects\cite{MSW1,MSW2,MSW3,MSW4}. After travelling the cosmic
distance to reach Earth, the arriving neutrinos are mass
eigenstates, which then oscillate in flavors while going through
Earth matter. Therefore, we also have to consider Earth matter
effects\cite{EM1,EM2,EM3} when we compute the event numbers of the
various flavors of neutrions.

Let $P_{\nu\nu}$ represent the collective effects of
neutrino-neutrino interactions which is a stepwise flavor conversion
probability of neutrinos at a critical energy $E_C$. In order to
obtain a simple expression for $P_{\nu\nu}$ for the neutrino and
$\bar{P}_{\nu\nu}$ for the antineutrino, we take a constant matter
density and box-spectra for both the neutrino and
antineutrino\cite{Collective5,Collective6,Collective7}. An analysis
of the collective effects in the case of three flavors has been made
in Ref.~\citep{Collective8}, and it allows us to characterize the
collective oscillation effects and to write down the flavor spectra
of the neutrino and antineutrino arriving at Earth. Following
Ref.~\citep{Collective8}, we have $P_{\nu\nu}=\bar{P}_{\nu\nu}=1$
for the normal hierarchy; and $\bar{P}_{\nu\nu}=1$, while
\begin{equation}
P_{\nu\nu}=
\begin{cases} 1 & (E<E_C), \\
 0 & (E>E_C),
 \end{cases}
\label{Pnunu}
\end{equation}
for the inverted hierarchy, where
$E_C=7MeV$\cite{Collective8,Collective9}.

There are two MSW resonance regions: the high resonance region and
the low resonance region. Let us denote the probability that the
neutrinos jump from one mass eigenstate to another at the high (low)
resonance layer by $P_H$ ($P_L$). Using the Landau-Zener
formula\cite{Landau1,Landau2}, one can obtain that
\begin{eqnarray}
P_H=\frac{\exp(-\frac{\pi}{2}\gamma
F)-\exp[-\frac{\pi}{2}\gamma(\frac{F}{\sin^2\theta_{13}})]}
{1-\exp[-\frac{\pi}{2}\gamma(\frac{F}{\sin^2\theta_{13}})]}, \label{GLPH}\\
\gamma=\frac{|\Delta m_{13}^2|
}{2E}\frac{\sin^22\theta_{13}}{\cos2\theta_{13}}\frac{1}{|d\ln
N_e/dr|_{res}}, \nonumber
\end{eqnarray}
where $N_e$ is the electron density and $F$ can be calculated by
Landau's method\cite{PH1}. Using the SN matter density profile
$\rho\approx C\cdot({10^7cm}/{r})^3\cdot10^{10} g/cm^3$ where C is a
structure constant between 1 and 15 \cite{SN3,SN4,SN5} and
$F\approx1$ (small $\theta_{13}$), we can obtain an simple
expression of $P_H$ that\cite{EM3,PH1,PH2,PH3}
\begin{eqnarray}
P_H=\exp\Bigg\{-\frac{\pi}{12}\Bigg[\frac{10^{10}
MeV}{E}\Bigg(\frac{\sin^32\theta_{13}}{\cos^22\theta_{13}}\Bigg)\times \nonumber\\
 \Bigg(\frac{|\Delta m^2_{32}|}{1 eV^2}\Bigg)C^{1/2}\Bigg]^{2/3}\Bigg\}, \label{PH}
\end{eqnarray}
where $|\Delta m^2_{32}|=2.6\times10^{-3}eV^2$. Similarity, we can
calculate the expression of the crossing probability at the low
resonance region inside the SN, $P_L$. However, due to the large
angle solution of the neutrino mixing, $P_L\approx0$.

\begin{center}
\includegraphics[width=0.35\textwidth]{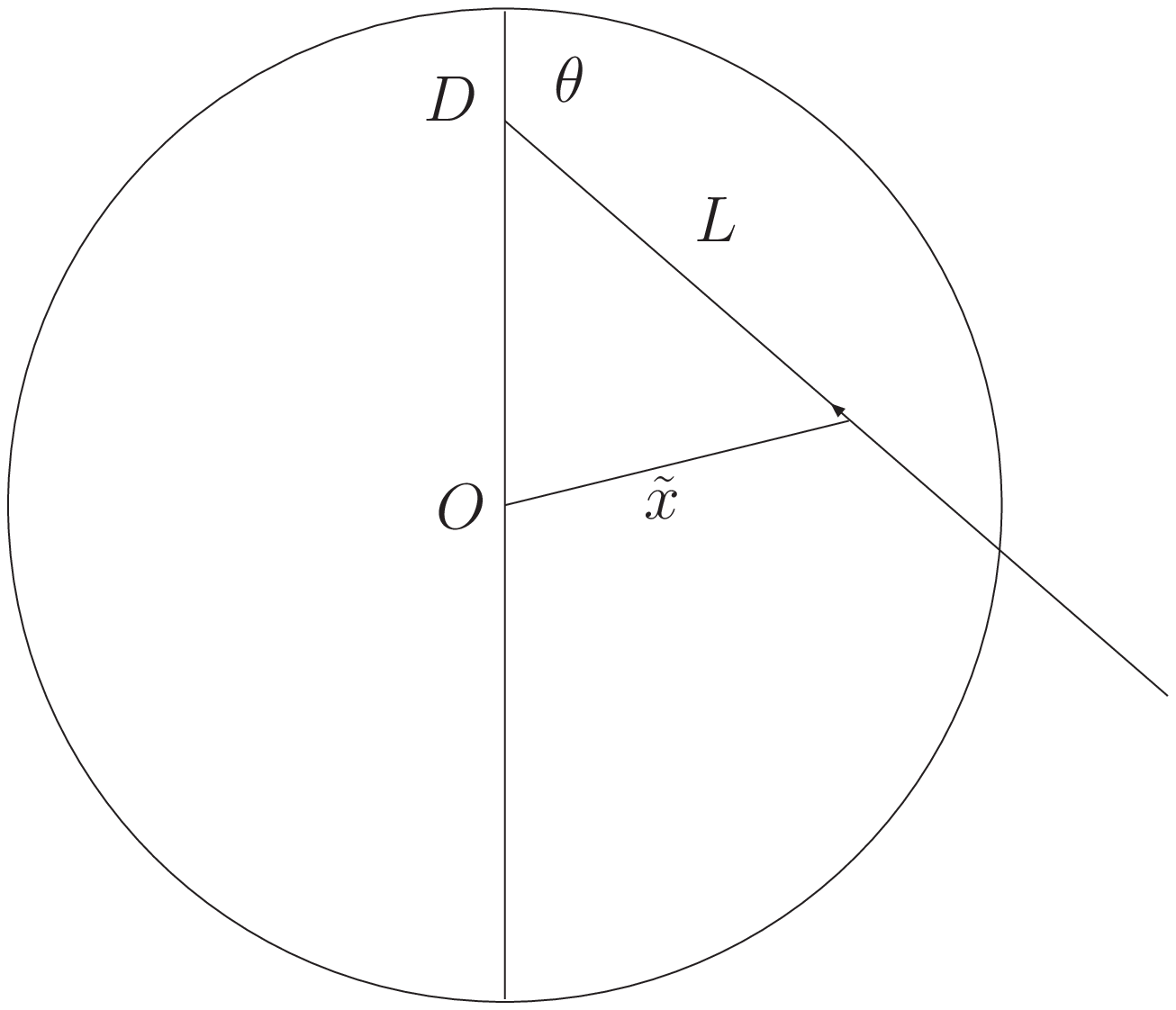}
\figcaption{\label{fig1} Illustration of the path of the SN neutrino
reaching the detector underground in Earth. $D$ is the location of
the detector, $\theta$ is the incident angle of the neutrino, $O$ is
the center of Earth, $L$ is the distance the neutrino travels
through Earth, and $\tilde{x}$ is the distance of the neutrino to
the center of Earth.}
\end{center}

Suppose a neutrino reaches the detector at the incident angle
$\theta$ (see Fig. 1). Then the distance that neutrino travels
through Earth is
\begin{equation}
 L=(-R_E+h)\cos{\theta}+\sqrt{R_E^2-(R_E-h)^2\sin^2{\theta}}, \label{L}
\end{equation}
where $h$ ($\approx0.4km$ for the Daya Bay experiment) is the depth
of the detector in and $R_E = 6400~{\rm km}$ is the radius of Earth.
Let $x$ be the distance that the neutrino travels into Earth, then
the distance of the neutrino to the center of Earth, $\tilde{x}$, is
given by
\begin{equation}
 \tilde{x}=\sqrt{(-R_E+h)^2+(L-x)^2+2(R_E-h)(L-x)\cos{\theta}}.
 \nonumber
\end{equation}
Let $P_{ie}$ be the probability that a neutrino mass eigenstate
$\nu_i$ enters the surface of Earth and arrives at the detector as
an electron neutrino $\nu_e$, one obtains\cite{EM2,EM3}
\begin{equation}
P_{2e}=\sin^2\theta_{12}+\frac{1}{2}\sin^22{\theta_{12}}\int_{x_0}^{x_f}dxV(x)\sin\phi
_{x\rightarrow x_f}^{m}, \label{P2e}
\end{equation}
where $\theta_{12}=32.5^{\circ}$, the potential $V(x)$ that the
electron neutrino experiences in Earth is
$\sqrt{2}G_F\rho(x)/(m_p+m_n)$ and $\phi_{a\rightarrow b}^m$ is
defined as
\begin{eqnarray}
\phi_{a\rightarrow b}^m &=& \int_a^b {\rm d}x \triangle_m(x),\nonumber \\
\triangle_m(x)&=&\frac{\triangle
 m_{21}^2}{2E}\sqrt{(\cos2\theta_{12}-\varepsilon(x))^2+\sin^22\theta_{12}},
 \nonumber
\end{eqnarray}
where $\varepsilon(x)={2EV(x)}/{\triangle m_{21}^2}$ and $\rho(x)$
is the matter mass density inside Earth\cite{Earth}.

In the following, we calculate the event numbers $N(i)$ of SN
neutrinos that can be observed through various reaction channels
"$i$". This is done by integrating over the neutrino energy $E$ of
the product of the target number $N_T$, the cross section of each
channel $\sigma(i)$, and the neutrino flux function at the
detector $F_{\alpha}^D(E)/4\pi D^2$,
\begin{equation}
 N(i)=N_T\int{{\rm d}E\cdot\sigma(i)\cdot\frac{1}{4\pi
 D^2}\cdot F_{\alpha}^D}, \label{Ntotal}
\end{equation}
where $\alpha$ stands for the neutrino or antineutrino of a given
flavor, and $D$ ($10~ kpc$ in the present discussion) is the
distance between the SN and Earth. After a straightforward
calculation, the fluxes at the detector can be
obtained\cite{Collective8,Collective9} ($x=\mu,\tau$):
\begin{eqnarray}
F_{\nu_e}^{D(N)}&=&P_{2e}P_HF_{\nu_e}^{(0)}+(1-P_{2e}P_H)F_{\nu_x}^{(0)},
\nonumber \\
F_{\bar{\nu}_e}^{D(N)}&=&(1-P_{2e})F_{\bar{\nu}_e}^{(0)}+P_{2e}F_{\bar{\nu}_x}^{(0)},
\nonumber \\
2F_{\nu_x}^{D(N)}&=&(1-P_{2e}P_H)F_{\nu_e}^{(0)}+(1+P_{2e}P_H)F_{\nu_x}^{(0)},
\nonumber \\
2F_{\bar{\nu}_x}^{D(N)}&=&P_{2e}F_{\bar{\nu}_e}^{(0)}+(2-P_{2e})F_{\bar{\nu}_x}^{(0)},
\label{FDN}
\end{eqnarray}
for the normal hierarchy ($\triangle m_{31}^2>0$), and
\begin{eqnarray}
F_{\nu_e}^{D(I)}&=&
\begin{cases} P_{2e}F_{\nu_e}^{(0)}+(1-P_{2e})F_{\nu_x}^{(0)}, & (E<E_C) \\
 F_{\nu_x}^{(0)}, & (E>E_C)
 \end{cases}
\nonumber \\
F_{\bar{\nu}_e}^{D(I)}&=&\bar{P}_H(1-\bar{P}_{2e})F_{\bar{\nu}_e}^{(0)}+(1+
\bar{P}_{2e}\bar{P}_H-\bar{P}_H)F_{\bar{\nu}_x}^{(0)},
\nonumber \\
2F_{\nu_x}^{D(I)}&=&
\begin{cases} (1-P_{2e})F_{\nu_e}^{(0)}+(1+P_{2e})F_{\nu_x}^{(0)}, & (E<E_C) \\
 F_{\nu_e}^{(0)}+F_{\nu_x}^{(0)}, & (E>E_C)
 \end{cases}
\nonumber \\
2F_{\bar{\nu}_x}^{D(I)}&=&(1+P_{2e}P_H-P_H)F_{\bar{\nu}_e}^{(0)} \nonumber\\
         &&+(1+P_H-P_{2e}P_H)F_{\bar{\nu}_x}^{(0)},\label{FDI}
\end{eqnarray}
for the inverted hierarchy ($\triangle m_{31}^2<0$). In Eqs.
(\ref{FDN}) and (\ref{FDI}), $F^{(0)}_{\nu_{\alpha}}$ is the
time-integrated neutrino energy spectrum of flavor $\alpha$ in
vacuum which can be described by the Fermi-Dirac distribution
\begin{equation}
 F_{\alpha}^{(0)}(E)=\frac{L_\alpha^{(0)}}{F_{\alpha
 3}T_{\alpha}^{4}}\frac{E^2}{\exp{(E/T_{\alpha}-\eta_\alpha)}+1},
 \label{Foa}
\end{equation}
where $T_{\alpha}$ is the temperature of the
neutrino\cite{Janka1,Janka2}
\begin{eqnarray}
 T_{\nu_e}=3-4MeV, && T_{\bar{\nu}_e}=5-6MeV, \nonumber  \\
 T_{\nu_x}&=&7-9MeV, \label{T}
 \end{eqnarray}
$\eta_\alpha$ is the pinching parameter of the spectra to represent
the deviation from being exactly thermal\cite{Janka1,Janka2}
\begin{equation}
 \eta_{\nu_e}\approx3-5, \quad \eta_{\bar{\nu}_e}\approx2.0-2.5,\quad
 \eta_{\nu_x}\approx0-2, \label{eta}
 \end{equation}
 $L^{(0)}_{\alpha}$ is the the luminosity, and $F_{\alpha j}$ is
defined by
\begin{equation}
 F_{\alpha j}=\int_{0}^{\infty}\frac{x^j}{\exp{(x-\eta_\alpha)}+1}{\rm
 d}x. \nonumber
\end{equation}

In the next section, using the relation between the event number of
SN neutrinos detected at Daya Bay, $N$, and the mixing angle
$\theta_{13}$, we will propose a possible method to acquire
information about $\theta_{13}$ smaller than $1.5^\circ$.

\section{\label{sec:hamckm}Acquire information about $\theta_{13}$
smaller than $1.5^\circ$ at Daya Bay}

It can be seen from the above section that, using Eqs.
(\ref{Pnunu}), (\ref{PH}), (\ref{P2e}), (\ref{Ntotal}),
(\ref{FDN}), and (\ref{FDI}), we can obtain the relation between
the event number of different flavor SN neutrinos, N, and the
mixing angle $\theta_{13}$. Let $R$ be the ratio of the event
number of $\nu_e$ over that of $\bar{\nu}_e$, one can obtain $R$
as a function of $\theta_{13}$. Therefore, we can propose a
possible method to acquire some information about the mixing angle
$\theta_{13}$ by detecting SN neutrinos. In this section, we will
just consider the process of the neutrino-carbon reactions, since
our method is not suitable for the inverse beta-decay and the
neutrino-electron reactions in the Daya Bay experiment. The
details are discussed at length in Ref.~\citep{EM3}.

The Daya Bay Collaboration uses LAB as the main part of the liquid
scintillator and the total detector mass is about 300 tons. LAB
has a chemical composition including $C$ and $H$. In our
calculation, the ratio of the numbers of $C$ and $H$, $N_C/N_H$,
is about 0.6. Therefore, the total numbers of target protons
$N_T^{(C)}=1.32\times10^{31}$. For the neutrino-neutrino
reactions, the effective cross sections are as follows\cite{Chen}:
\begin{eqnarray}
 &&\ \langle\sigma(^{12}C(\nu_e,e^{-})^{12}N)\rangle=1.85\times10^{-43}
cm^2,\nonumber \\
  &&\ \langle\sigma(^{12}C(\bar{\nu}_e,e^{+})^{12}B)\rangle=1.87\times10^{-42}
cm^2, \label{sigma CB}
\end{eqnarray}
for the charged-current capture,  and
\begin{eqnarray}
 &&\ \langle\sigma(\nu_e^{12}C)\rangle=1.33\times10^{-43}cm^2, \nonumber \\
 &&\ \langle\sigma(\Bar{\nu}_e^{12}C)\rangle=6.88\times10^{-43}cm^2,
 \label{sigma CC} \\
 &&\
 \langle\sigma(\nu_x(\Bar{\nu}_x)^{12}C)\rangle=3.73\times10^{-42}cm^2, \quad
 x=\mu,\tau,
 \nonumber
\end{eqnarray}
for the neutral-current capture. Using Eqs. (\ref{Pnunu}),
(\ref{PH}), (\ref{P2e}), (\ref{Ntotal}), (\ref{FDN}), (\ref{FDI}),
(\ref{sigma CB}) and (\ref{sigma CC}), we plot the ratio $R$ of
the event number of $\nu_e$ over that of $\bar{\nu}_e$ which can
be detected at Daya Bay as a function of the mixing angle
$\theta_{13}$. The result is shown in Fig. 2.
\begin{center}
\includegraphics[width=0.5\textwidth]{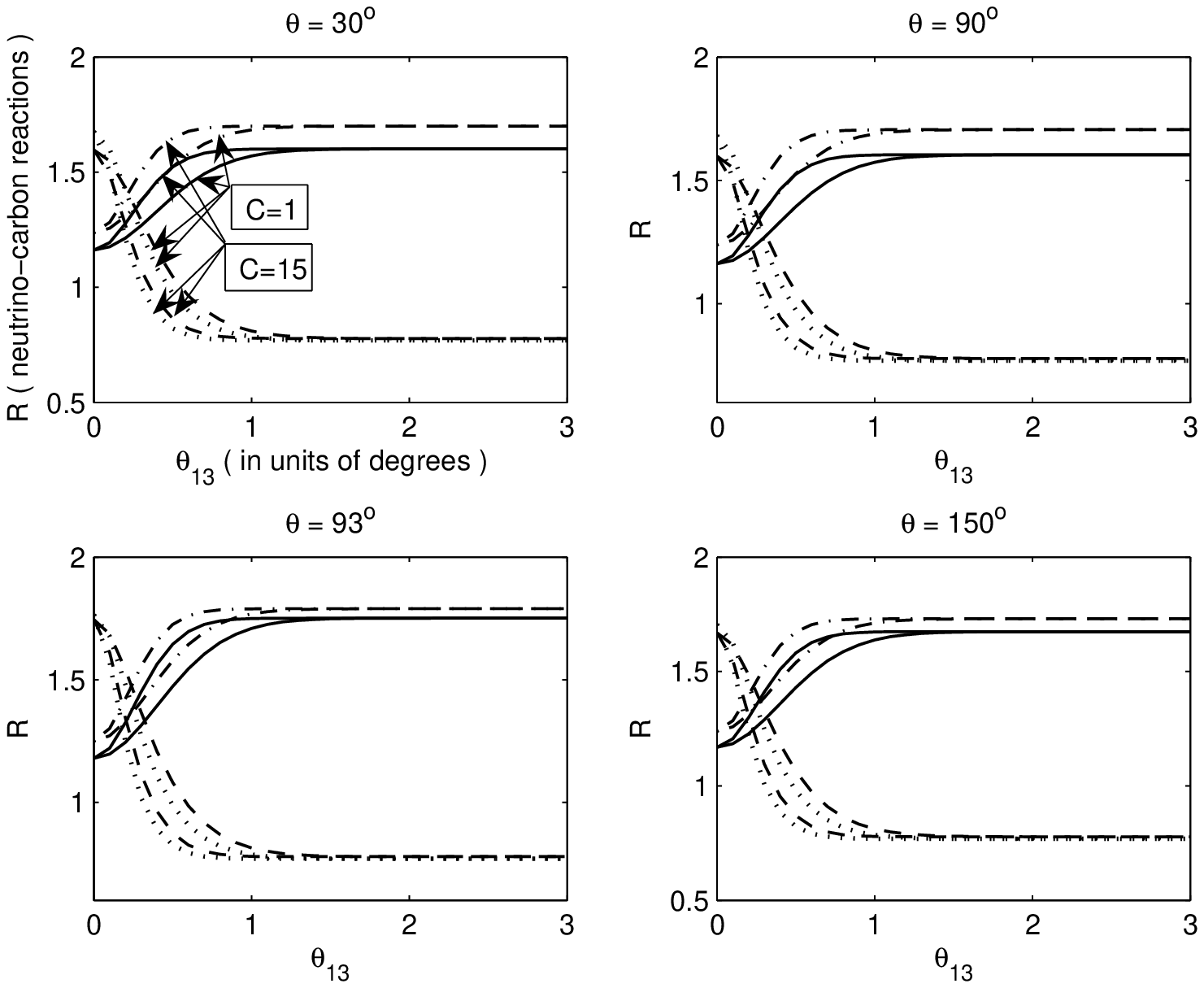}
\figcaption{\label{fig2} The ratio of the event number of $\nu_e$
to that of $\bar{\nu}_e$, $R$, as a function of the mixing angle
$\theta_{13}$ in the channel of neutrino-carbon reactions at the
Daya Bay experiment. The incident angle is (a) $\theta=30^\circ$;
(b) $\theta=90^\circ$; (c) $\theta=93^\circ$; (d)
$\theta=150^\circ$. The solid curves correspond to the normal
hierarchy (max), the dashed curves correspond to the inverted
hierarchy (max), the dot-dashed curves correspond to the normal
hierarchy (min), the dotted curves correspond to the inverted
hierarchy (min), where "max" ("min") corresponds to the maximum
(minimum) values of $T_{\alpha}$ and $\eta_{\alpha}$.}
\end{center}

It can be seen from Fig. 2 that the uncertainties of $R$ due to
$T_{\alpha}$ and $\eta_{\alpha}$ are not large. For
$\theta_{13}\leq1.5^\circ$, $R$ is very sensitive to $\theta_{13}$.
However, while $\theta_{13}>1.5^\circ$, $R$ is nearly independent of
$\theta_{13}$. Therefore, when $\theta_{13}$ is smaller than
$1.5^{\circ}$, we may restrict the mixing angle $\theta_{13}$ in a
small range and get information about mass hierarchy by detecting
the ratio of event numbers of SN neutrinos even though there are
still some uncertainties due to the incident angle $\theta$, the
mass hierarchy $\triangle m_{31}^2$, and the structure coefficient
$C$ of the SN density function.

At the Daya Bay experiment, the sensitivity of
$\sin^22\theta_{13}$ will reach 0.01, i.e., to determine
$\theta_{13}$ down to about $3^\circ$. Therefore, if the actual
value of $\theta_{13}$ is smaller than $3^\circ$, the Daya Bay
experiment can only provide an upper limit for $\theta_{13}$.
However, if an SN explosion takes place during the operation of
Daya Bay, roughly within the cosmic distance considered here, it
is possible to reach a much smaller value of $\theta_{13}$ through
the ratio of the event numbers of different flavor SN neutrinos in
the channel of neutrino-carbon reactions as discussed above. It is
interesting to note that because of the multi detectors set up,
experiments such as the Daya Bay have an internal coincidence
check for SN neutrino events.

\section{\label{sec:hamckm}Summary and discussion}
In this paper, we first discuss the detection of SN neutrinos on
Earth. Since neutrino flavor conversions inside the SN depend on
the neutrino mixing angle $\theta_{13}$, we give a possible method
to acquire information about $\theta_{13}$ smaller than
$1.5^{\circ}$ by detecting SN neutrinos at Daya Bay.

We let the parameters in the neutrino energy spectra (the
temperatures and the pinching parameters) vary in some reasonable
ranges. In fact, the simulations from the two leading groups, the
Livermore group\cite{Totani} and the Garching
group\cite{Keil1,Keil2}, led to parameters which agree within about
20-30\%. However, their central values of SN parameters are
different.

\end{multicols}

\vspace{-2mm}
\centerline{\rule{80mm}{0.1pt}}
\vspace{2mm}

\begin{multicols}{2}

\end{multicols}


\begin{thebibliography}{90}

\vspace{3mm}

\bibitem{SN1} Kotake K, Sato K, Takahashi K, Rept. Prog.
Phys., 2006, {\bf 69}: 971---1143
\bibitem{SN2} Dighe A S, {\it Physics potential of future supernova
neutrino observation}, arXiv:hep-ph/0809.2977, {\it Neutrinos from
a core collapse supernova}, arXiv:hep-ph/0712.4386
\bibitem{DayaBay} [Daya Bay] GUO X H et al., {\it A
precision measurement of the neutrino mixing angle $\theta_{13}$
using reactor antineutrinos at Daya Bay}, arXiv:hep-ex/0701029

\bibitem{Collective1} DUAN H Y, Fuller G M, QIAN Y Z, Phys. Rev.
D, 2006, {\bf 74}: 123004
\bibitem{Collective2} DUAN H Y, Fuller G M, Carlson J et al.
Phys. Rev. D, 2006, {\bf 74}: 105014
\bibitem{Collective3} DUAN H Y, Fuller G M, Carlson J et al.
Phys. Rev. D, 2007, {\bf 75}: 125005
\bibitem{Collective4} DUAN H Y, Fuller G M, Carlson J, Comp. Scie. $\&$ Disc.,
2008, {\bf 1}: 015007
\bibitem{Collective5} Hannestad S, Raffelt G G, Sigl G et al. Phys. Rev.
D, 2006, {\bf 74}: 105010
\bibitem{Collective6} Raffelt G G, Smirnov A Y, Phys. Rev. D,
2007, {\bf 76}: 081301
\bibitem{Collective7} Raffelt G G, Smirnov A Y, Phys. Rev. D, 2007, {\bf 76}:
125008
\bibitem{Collective8} Dusgupta B, Dighe A, Phys. Rev. D, 2008, {\bf
77}: 113002
\bibitem{Collective9} Chakraborty S, Choubey S, Dasgupta B et al. J. Cosmol.
Astropart. Phys., 2008, {\bf 09}: 013

\bibitem{MSW1} Wolfenstein L, Phys. Rev. D, 1978, {\bf 17}: 2369---2374
\bibitem{MSW2} Wolfenstein L, Phys. Rev. D, 1979, {\bf 20}: 2634---2375
\bibitem{MSW3} Mikheyev S P, Smirnov A Y, Sov. J. Nucl. Phys.,
1985, {\bf 42}: 913---917
\bibitem{MSW4} Mikheyev S P, Smirnov A Y, Nuovo Cimento C, 1986, {\bf 9}:
17---26
\bibitem{EM1} Dighe A S, Smirnov A Y, Phys. Rev. D, 2000, {\bf 62}: 033007
\bibitem{EM2} GUO X H, YOUNG B L, Phys. Rev. D, 2006, {\bf 73}: 093003
\bibitem{EM3} GUO X H, HUANG M Y, YOUNG B L, {\it Realistic Earth
matter effects and a method to measure small $\theta_{13}$ in the
detection of supernova neutrinos}, arXiv:0806.2720 [hep-ph]

\bibitem{Landau1} Zener C, Proc. R. Soc. London A, 1932, {\bf 137}:
696---702
\bibitem{Landau2} Landau L D, Lifshitz E M. Quantum Mechanics:
Non-relativistic Theory. New York: Pergamon, 1977. 199---237
\bibitem{PH1} Kuo T K, Pantaleone J, Rev. Mod. Phys., 1989, {\bf
61}: 937---979
\bibitem{PH2} Lunardini C, Smirnov A Y, J. Cosmol. Astropart.
Phys., 2003, {\bf 06}: 009
\bibitem{PH3} Kachelrie$\beta$ M, Strumia A,
Tom$\grave{a}$s R et al. Phys. Rev. D, 2002, {\bf 65}: 073016
\bibitem{book1} Giunti C, Kim C W. Fundamentals of Neutrino Physics
and Astrophysics. New York: Oxford, 2007. 511---539
\bibitem{book2} Shapiro S L, Teukolsky S A. Black Holes, White
Dwarfs, and Neutron Stars. New York: Wiley, 1983. 512---542
\bibitem{book3} Mohapatra R N, Pal P B. Massive Neutrino in Physics
and Astrophysics, 3nd ed. Singapore: World Scientific, 2004.
340---358
 \bibitem{SN3} Brown G E, Bethe H A, Baym G, Nucl. Phys. A,
 1982, {\bf 375}: 481---532
 \bibitem{SN4} Bethe H A, Rev. Mod. Phys., 1990, {\bf
62}: 801---866
 \bibitem{SN5} Janka H T, Astron. Astrophys., 2001, {\bf 368}:
527---560
\bibitem{Earth} Dziewonski A M, Anderson D L, Phys. Earth Planetary
Interiors, 1981, {\bf 25}: 297---356
\bibitem{Janka1} Janka H T, Hillebrandt W, Astron. Astrophys., 1989,
{\bf 224}: 49---56
 \bibitem{Janka2} Janka H T, Hillebrandt W, Astron. Astrophys. Suppl.
Ser., 1989, {\bf 78}: 375---397
\bibitem{Chen} Cadonati L, Calaprice F P, Chen M C, Astropart.
Phys., 2002 {\bf 16}, 361-372
\bibitem{Totani} Totani T, Sato K, Dalhed H E et al. Astrophys.
J., 1998, {\bf 496}: 216---225
\bibitem{Keil1} Keil M T, Raffelt G G, Janka H T, Astrophys. J., 2003,
{\bf 590}: 971---991
\bibitem{Keil2} Keil M T, {\it Supernova neutrino spectra and
applications to flavor oscillations}, arXiv:astro-ph/0308228
\end{thebibliography}
\end{document}